\documentclass[superscriptaddress,prl,twocolumn]{revtex4}
\usepackage{amsmath}
\usepackage{amssymb}
\usepackage{graphicx}
\usepackage{bm}
\usepackage{dcolumn}
\usepackage{color}

\begin{document}

\title{Slow quantum oscillations without fine-grained Fermi Surface
Reconstruction in Cuprate Superconductors}
\author{P.D. Grigoriev}
\affiliation{L. D. Landau Institute for Theoretical Physics, 142432 Chernogolovka, Russia}
\affiliation{National University of Science and Technology ``MISiS'', Moscow 119049,
Russia}
\affiliation{P.N. Lebedev Physical Institute, RAS, 119991, Moscow, Russia}
\author{Timothy Ziman}
\affiliation{Institut Laue-Langevin, BP 156, 41 Avenue des Martyrs, 38042 Grenoble Cedex
9, France}
\affiliation{LPMMC (UMR 5493), Universit\'e de Grenobles-Alpes and CNRS, Maison des
Magist\`eres, BP 166, 38042 Grenoble Cedex 9, France}

\begin{abstract}
The Fourier transform of the observed magnetic quantum oscillations (MQO) in
YBa$_{2}$Cu$_{3}$O$_{6+\delta }$ high-temperature superconductors has a
prominent low-frequency peak with two smaller neighbouring peaks.
The separation and positions of these three peaks are almost independent of doping.
This pattern has been explained previously by rather special, exquisitely
detailed, Fermi-surface reconstruction. We propose that these MQO have a
different origin, and their frequencies are related to the bilayer and
inter-bilayer electron hopping rather than directly to the areas of tiny
Fermi-surface pockets. Such so-called ``slow oscillations" explain more
naturally many features of the observed oscillations and allow us to
estimate the inter-layer transfer integrals and in-plane Fermi momentum.
\end{abstract}

\date{\today }
\pacs{74.72.-h,72.15.Gd,73.43.Qt,74.70.-b,71.45.Lr}
\maketitle


Magnetic quantum oscillations (MQO) provide a traditional and powerful tool
to study the Fermi surface and other electronic structure parameters of
various metals.\cite{Abrik,Shoenberg,Ziman} In the last decade, following
their first observation in cuprate high-temperature superconductors,\cite%
{ProustNature2007} they have been extensively used to investigate the
electronic structure of cuprates, both hole-\cite{AnnuReviewYBCO2015} and
electron-doped,\cite{HelmNd2015} as well as in Fe-based superconductors.\cite%
{BaFeAs2011,Graf2012,ColdeaReview2013,FeSeTerashima2014,FeSeAudouard2015,FeSeWatsonPRB2015,FeSeMQOPRL2015}
Probably most 
surprising are the data for the underdoped yttrium barium copper oxide
(YBCO) compounds YBa$_{2}$Cu$_{3}$O$_{6+\delta }$ (\cite%
{ProustNature2007,SebastianNature2008,AudouardPRL2009,SingletonPRL2010,SebastianPNAS2010,SebastianPRB2010,SebastianPRL2012,SebastianNature2014,ProustNatureComm2015}
reviewed in \cite%
{SebastianRepProgPhys2012,ProustComptesRendus2013,AnnuReviewYBCO2015,SebastianPhilTrans2011}%
). The Fourier transform of these quantum oscillations has a prominent peak
at frequency $F_{\alpha }\approx 530T$ with two smaller shoulders at $F_{\pm
}=F_{\alpha }\pm \Delta F_{\alpha}$, where $\Delta F_{\alpha}\approx 90T$.
All these frequencies are much smaller than expected from closed pockets of
the Fermi surface(FS). Many different theoretical models have been proposed
to explain such a set of frequencies (\cite%
{EfimovPRB2008,GarciaNJP2010,HarrisonNJP2012,HarrisonSciRep2015,Briffa2015},
reviewed in \cite%
{SebastianRepProgPhys2012,ProustComptesRendus2013,AnnuReviewYBCO2015}).
While these interpretations vary in detail, such as inclusion of spin-orbit
or Zeeman splittings, they are all based on Fermi-surface reconstruction due
to the periodic potential created by a charge density wave (CDW). A weak,
probably, inhomogeneous or fluctuating CDW order has been detected in YBa$%
_{2}$Cu$_{3}$O$_{6+\delta }$ compounds by X-Ray scattering \cite%
{XRayScience2012,XRayNatPhys2012,XRayPRL2012,Xray2016}, nuclear magnetic
resonance \cite{NMR2011Wu,NMR2015Wu}, and sound velocity measurements \cite%
{CDWSoundVelocity}. High magnetic fields suppress superconductivity and lead
to long-range CDW coherence.\cite{Xray2016} Static CDW order can indeed lead
to Fermi surface reconstruction, seen in new MQO frequencies,
but only if the 
CDW potential is 
sufficiently strong. More precisely, the CDW energy gap must be larger than
the magnetic-breakdown gap $\Delta _{MB}\sim \sqrt{\hbar \omega _{c}E_{F}}$,%
\cite{Shoenberg} where $\hbar \omega _{c}$ is the cyclotron energy, \textit{%
i.e.} the separation between the Landau levels, and $E_{F}\sim 1$eV is the
Fermi energy of the unreconstructed electron dispersion. The oscillations in
cuprates are measured in magnetic fields $\boldsymbol{B}$ higher than $30$
tesla, where $\Delta _{MB}\gtrsim 40$meV is rather large and a fluctuating
CDW ordering may not be enough to form new frequencies with amplitudes
sufficient for experimental observation. Note that a frequency pattern
somewhat similar to that of YBa$_{2}$Cu$_{3}$O$_{6+\delta } $ is observed in
the closely related stoichiometric compound YBa$_{2}$Cu$_{4} $O$_{8}$,\cite%
{Yelland2008,Bangura2008,TanPNAS2015} where there is no experimental
indication of a static superstructure. Even if this CDW is sufficiently
strong, it is hard to explain the observed three-peak frequency pattern of
MQO in YBCO without additional frequencies of similar amplitude from the CDW
wave vector seen in X-ray experiments. \cite%
{XRayScience2012,XRayNatPhys2012,XRayPRL2012,Xray2016} Moreover, if FS
reconstruction really is the origin of the observed $F_{\alpha }$, $F_{+}$,
and $F_{-}$, they should depend strongly on doping.\cite{CommentDoping1}
Such a strong dependence 
is, in fact, observed in the electron-doped cuprate superconductor Nd$_{2-x}$%
Ce$_{x}$CuO$_{4}$,\cite{HelmNd2009,HelmNd2010} where the frequency changes
by 20\%, from $290$ to $245$ T, when the doping level changes from 0.15 to
0.17. In contrast, in YBa$_{2}$Cu$_{3}$O$_{6+\delta }$ the observed doping
dependence of the three low frequencies is much weaker. The data are
somewhat controversial: frequency changes vary from very weak, less than 5\%
(Ref. \cite{SingletonPRL2010}, Fig.4) to about 25\% (Ref. \cite%
{ProustComptesRendus2013}, Fig. 5b ) when the doping changes from $p\approx
0.1$ to 0.14. In a recent study $F_{\alpha }$ changes by only 10\%, from
515T to 570T, for $p$ varying from 0.09 to 0.14.\cite{DopingDependence2015}
Within the CDW scenario, a model of compensated FS pockets has been proposed,%
\cite{SebastianPRB2010} but the doping-dependence remains problematic. To
avoid the problematic FS reconstruction scenarios an unusual source of
oscillations was proposed in terms of Andreev-type bound states,\cite%
{NPhysPereg} but the predicted change in oscillation frequencies with
superconducting gap contradicts experiment.
In this paper, we propose a simple alternative picture, which is consistent
with the observed three equidistant magnetic oscillation peaks in YBa$_{2}$Cu%
$_{3}$O$_{6+\delta }$ and other features of the measurements.

The unreconstructed FS of YBCO consists of one large pocket, almost a square
with smoothed corners, that fills about one half of the Brillouin zone and
would correspond to a large frequency $\sim 10^{4}$ tesla. The Fermi-surface
reconstruction, possibly caused by the pseudogap, AFM or CDW order, takes
place for doping level $p<10-15\% $, as suggested by ARPES \cite%
{FournierARPES2010}, or by observation of negative Hall \cite%
{ElPocketHall2007,BadouxHall2016} and Seebeck \cite{ElPocketSeebeck2010}
coefficients which indicate electron-like FS pockets. However, the observed
MQO frequency $F_{\alpha }\approx 530T$ corresponds to a FS cross-section of
only 2\% of the Brillouin zone, which is considerably less than the size of
the expected FS pockets even for a reconstructed FS. Analogous
\textquotedblleft slow oscillations \textquotedblright (SlO) have been
observed in organic superconductors and attributed not to new small FS
pockets, but to the interplay (mixing) of two close frequencies $F_{\beta
}\pm \Delta F$, where the frequency splitting $\Delta F$ is due to FS
warping, originating from the interlayer electron transfer integral $t_{z}$.%
\cite{SO} As a result, magnetoresistance (MR) oscillations with a new
frequency $F_{slow}=2\Delta F=4t_{z}B/\hbar \omega _{c}$ arise. The
amplitudes of the emergent slow oscillations may be even higher than those
of the oscillations with the original frequencies $F_{\beta }\pm \Delta F$,
because the SlO are damped neither by temperature nor by long-range disorder
(variations of $E_{F}$ along the sample on the scale much larger than
magnetic length).\cite{SO,Shub} The latter feature is especially important,
because even in high-quality monocrystals of organic metals this long-range
disorder makes the major contribution to the Dingle temperature.\cite{SO} In
the notoriously inhomogeneous cuprates, such disorder is much stronger than
in organic metals, and magnetic oscillations from closed pockets should be
strongly damped, whereas the proposed slow oscillations are insensitive to
this form of disorder and can easier be observed.

\begin{figure}[tbh]
\includegraphics[width=0.3\textwidth]{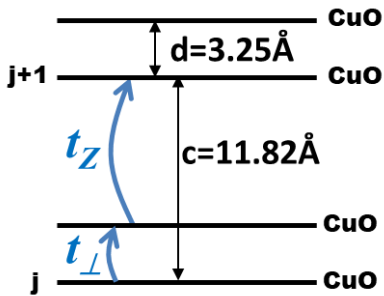}
\caption{Schematic illustration of the bilayer crystal structure in YBa$_{2}$%
Cu$_{3}$O$_{6+\protect\delta }$ and YBa$_{2}$Cu$_{4}$O$_{8}$
high-temperature superconductors with two interlayer transfer integrals $%
t_{\perp }$ and $t_{z}$.}
\label{FigBilayer}
\end{figure}

\begin{figure}[tbh]
a\includegraphics[width=0.21\textwidth]{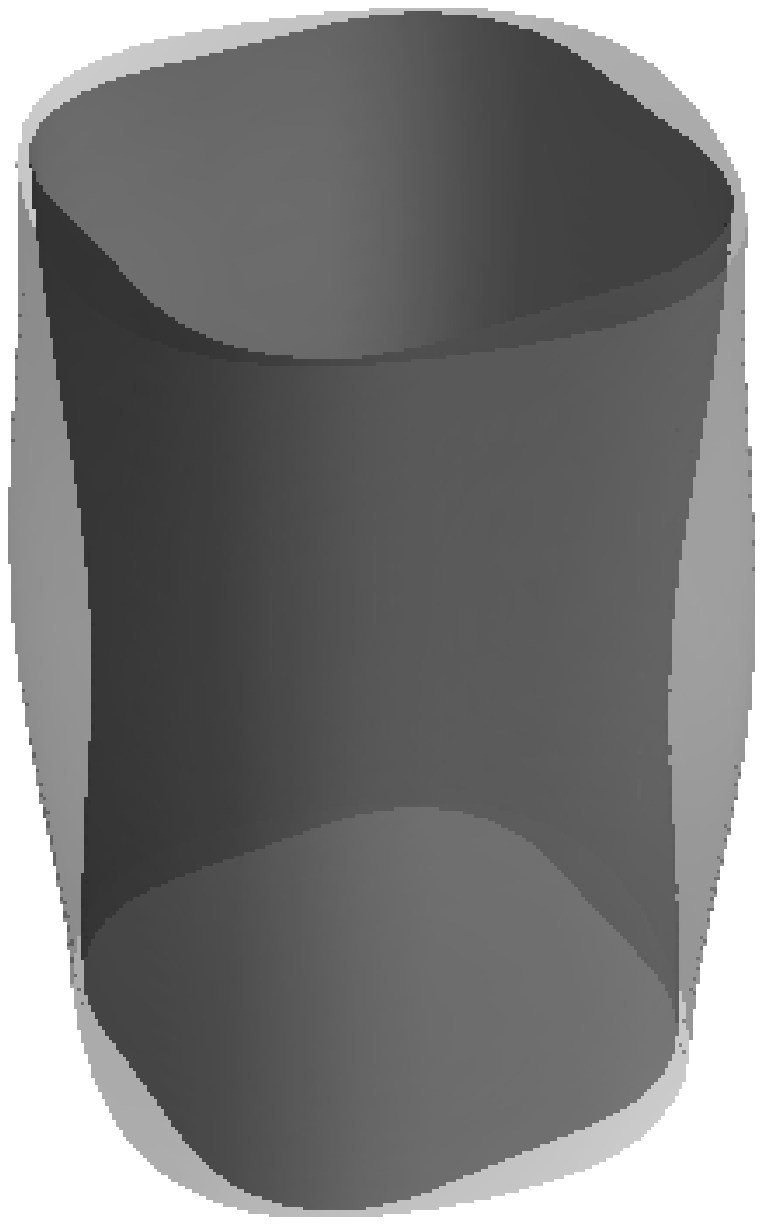} \ \ b%
\includegraphics[width=0.20\textwidth]{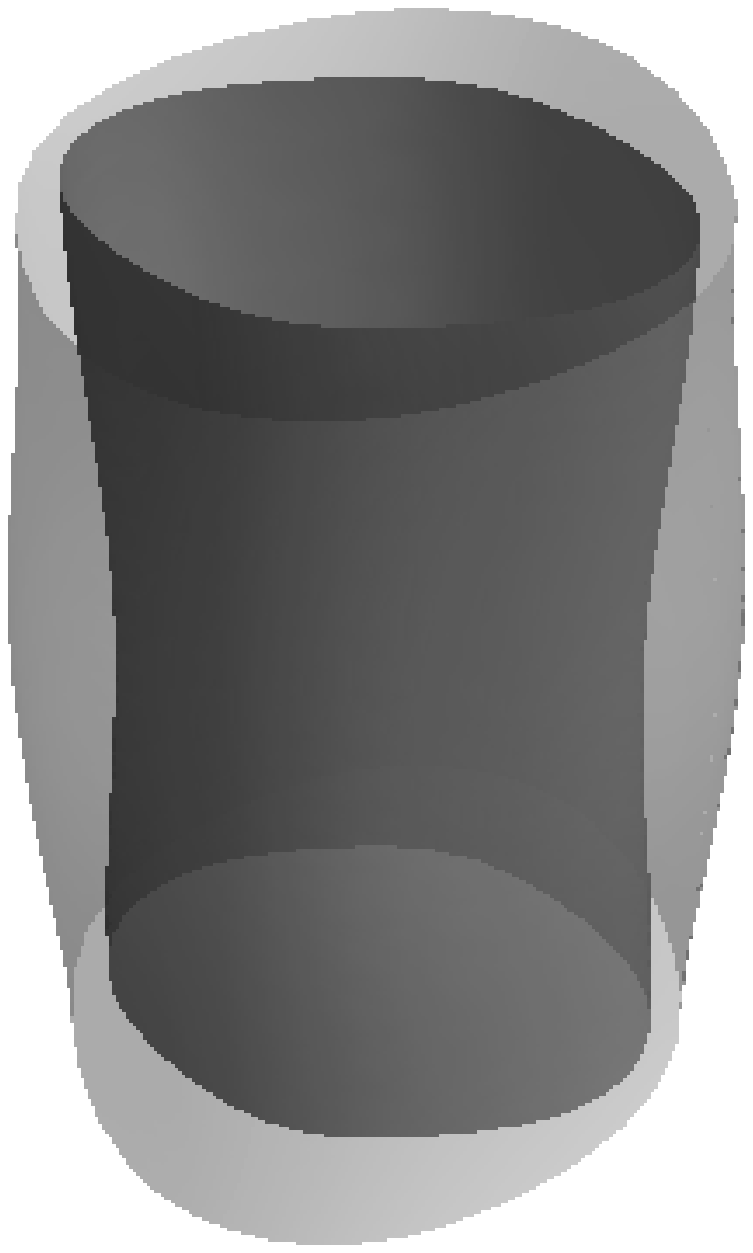}
\caption{A quasi-2D Fermi surface with interlayer warping due to $2t_{z}$
and double bilayer splitting, corresponding to dispersion in Eq. (\protect
\ref{EpsBilayerKzApprox}) for $t_{z}\left( \boldsymbol{k}_{\parallel
}\right) \approx const$, and (a) $t_{\perp }\left( \boldsymbol{k}_{\parallel
}\right) =t_{\perp }(1+0.3\sin 4\protect\phi )$ and (b) $t_{\perp }\left(
\boldsymbol{k}_{\parallel }\right) =t_{\perp }(1+0.5\sin 2\protect\phi )$,
where $\tan \protect\phi =k_y/k_x$.}
\label{FigFS}
\end{figure}

In contrast to the organic metal $\beta
$-(BEDT-TTF)$_{2}$IBr$_{2}$ with simple quasi-2D layered crystal
structure, where the SlO may originate only from the FS warping
due to $k_{z}$ electron dispersion\cite{SO}, the YBCO compounds
have bilayer crystal structures (see Fig. \ref{FigBilayer}). There
are then two types of interlayer hopping: (i) between adjacent
layers separated by distance $d$
within one bilayer, given by the transfer integral $t_{\perp }=t_{\perp }(%
\boldsymbol{k}_{\parallel })$, and (ii) between adjacent equivalent bilayers
separated by distance $h$, given by the transfer integral $t_{z}=t_{z}(%
\boldsymbol{k}_{\parallel })$, where $\boldsymbol{k}_{\parallel }$ is the
intralayer momentum. The resulting electron energy spectrum is given by \cite%
{GarciaNJP2010} (see, e.g., Eq. (6) of Ref. \onlinecite{GarciaNJP2010})
\begin{equation}
\epsilon _{\pm }\left( k_{z},\boldsymbol{k}_{\parallel }\right) =\epsilon
_{\parallel }\left( \boldsymbol{k}_{\parallel }\right) \pm \sqrt{%
t_{z}^{2}+t_{\perp }^{2}+2t_{z}t_{\perp }\cos \left[ k_{z}\left( h+d\right) %
\right] }.  \label{EpsBilayerKz}
\end{equation}%
For $t_{z}\ll t_{\perp }$ this spectrum contains bonding and antibonding
states separated by $\sim 2t_{\perp }\left( \boldsymbol{k}_{\parallel
}\right) $, each with weak $k_{z}$ dispersion:
\begin{equation}
\epsilon _{\pm }\left( k_{z},\boldsymbol{k}_{\parallel }\right) \approx
\epsilon _{\parallel }\left( \boldsymbol{k}_{\parallel }\right) \pm t_{\perp
}\left( \boldsymbol{k}_{\parallel }\right) \pm 2t_{z}\left( \boldsymbol{k}%
_{\parallel }\right) \cos \left[ k_{z}\left( h+d\right) \right] .
\label{EpsBilayerKzApprox}
\end{equation}%
The corresponding Fermi surface pocket is shown schematically in Fig. \ref%
{FigFS} for a slight tetragonal modulation of $\epsilon _{\parallel }\left(
\boldsymbol{k}_{\parallel }\right) $ and two different symmetries of $%
t_{\perp }\left( \boldsymbol{k}_{\parallel }\right) $. In YBCO there are
thus at least two types of splitting of the original frequencies: the larger
bilayer splitting $\Delta F_{\perp }=t_{\perp }B/\hbar \omega _{c}$, where $%
t_{\perp }=\left\langle t_{\perp }\left( \boldsymbol{k}_{\parallel }\right)
\right\rangle \neq 0$ and the angular brackets signify an averaging over
in-plane momentum $\boldsymbol{k}_{\parallel }$ on the FS, and the smaller
splitting $\Delta F_{c}=2t_{z}B/\hbar \omega _{c}\ll \Delta F_{\perp }\ll
F_{\beta }$ due to the $k_{z}$ electron dispersion, where we also assume $%
t_{z}=\left\langle t_{z}\left( \boldsymbol{k}_{\parallel }\right)
\right\rangle \neq 0$. These two splittings result in \textit{four}
underlying MQO frequencies $F_{\beta }\pm \Delta F_{\perp }\pm \Delta F_{c}$
of similar amplitudes, instead of only \textit{two} $F_{\beta }\pm \Delta
F_{c}$ for organic metals without bilayers\cite{SO}. The SlO of MR originate
from these \textit{four} frequencies in a similar way as before\cite{SO,RET}
but result in a much richer set of frequencies, as we show below.

At finite temperature the metallic conductivity along $i$-th axis $\sigma
_{i}=\sigma _{ii}$ is given by the sum of contributions from all ungapped FS
pockets $\beta $:%
\begin{equation}
\sigma _{i}=\sum_{\beta }\sigma _{i,\beta }=\sum_{\beta }e^{2}g_{F\beta
}D_{i,\beta }.  \label{s1}
\end{equation}%
Each pocket $\beta $ contributes to the total metallic conductivity along
axis $i$ at low temperature via the product of a density of electron states
(DoS) $g_{F,\beta }=g_{\beta }\left( \varepsilon =E_{F}\right) $ and an
electron diffusion coefficient $D_{i,\beta }$. Both contribute to
oscillations, since they vary with the magnetic field $B_{z}$ perpendicular
to the conducting $x$-$y$ layers as:
\begin{equation}
\frac{g_{F\beta }}{g_{0\beta }}=1+A_{\beta }\sum\limits_{j,l=\pm 1}\cos
\left( 2\pi \frac{F_{\beta }+j\Delta F_{\perp }+l\Delta F_{c}}{B_{z}}\right)
,  \label{rho0}
\end{equation}%
with $g_{0\beta }=m_{\beta }^{\ast }/\pi \hbar ^{2}d$ the average DoS at the
Fermi level and $m_{\beta }^{\ast }$ the effective mass for the pocket $%
\beta $, and
\begin{equation}
\frac{D_{i,\beta }}{D_{0i,\beta }}=1+B_{i,\beta }\sum\limits_{j,l=\pm 1}\cos
\left( 2\pi \frac{F_{\beta }+j\Delta F_{\perp }+l\Delta F_{c}}{B_{z}}\right)
,  \label{D0}
\end{equation}%
where the non-oscillating part $D_{0i,\beta }$ of the diffusion coefficient
in metals is proportional to the mean-square electron velocity $v_{i}$ at
the Fermi level. The harmonic amplitudes $A_{\beta }\sim B_{i,\beta }$
contain the Dingle factor\cite{Shoenberg,Dingle,Bychkov} $R_{D,\beta
}\approx \exp \left( -2\pi \Gamma /\hbar \omega _{c,\beta }\right) $, where
the electron level broadening $\Gamma =\Gamma \left( \boldsymbol{B},T\right)
$ comes from various types of interaction of the conducting electrons.\cite%
{CommentMB}

Combining Eqs. (\ref{s1})-(\ref{D0}) we obtain
\begin{eqnarray}
\frac{\sigma _{i,\beta }}{\sigma _{i,\beta }^{(0)}} &=&1+\left( A_{\beta
}+B_{i,\beta }\right) \sum\limits_{j,l=\pm 1}\cos \left( 2\pi \frac{F_{\beta
}+j\Delta F_{\perp }+l\Delta F_{c}}{B_{z}}\right)  \notag \\
&&+A_{\beta }B_{i,\beta }\sum\limits_{j,l,j^{\prime },l^{\prime }=\pm 1}\cos
\left( 2\pi \frac{F_{\beta }+j\Delta F_{\perp }+l\Delta F_{c}}{B_{z}}\right)
\times  \notag  \label{Slow0} \\
&&\times \cos \left( 2\pi \frac{F_{\beta }+j\,j^{\prime }\Delta F_{\perp
}+l\,l^{\prime }\Delta F_{c}}{B_{z}}\right) ,  \label{Slow1}
\end{eqnarray}%
where $\sigma _{i,\beta }^{(0)}=e^{2}g_{0\beta }D_{0i,\beta }$ does not
oscillate. The second term in Eq. (\ref{Slow1}) gives the usual MQO with
amplitudes $\sim A_{\beta }$ and four fundamental frequencies $F_{\beta }\pm
\Delta F_{\perp }\pm \Delta F_{c}\sim F_{\beta }\gg \Delta F_{\perp }$. The
last term in Eq. (\ref{Slow1}) is of the second order in the amplitude $%
A_{\beta }$ and gives various frequencies: (i) the 4 second harmonics $%
2\left( F_{\beta }\pm \Delta F_{\perp }\pm \Delta F_{c}\right) $, which are
strongly damped by temperature and disorder and can be neglected; (ii) for $%
j^{\prime }=1$ and $l^{\prime }=-1$ the SlO with very low frequency $2\Delta
F_{c}\ll \Delta F_{\perp }\ll F_{\beta }$; (iii) for $j^{\prime }=-1$ and $%
l^{\prime }=\pm 1$ the SlO with intermediate frequencies $2\Delta F_{\perp }$
and $2\Delta F_{\perp }\pm 2\Delta F_{c}$. Indeed,
neglecting the high frequency ($\sim 2F_{\beta }$) contributions
we can rewrite the last term in Eq. (\ref{Slow1}) for $j^{\prime }=-1$ as
\begin{gather}
\frac{A_{\beta }B_{i,\beta }}{2}\sum\limits_{j,l,l^{\prime }=\pm 1}\cos
\left( 2\pi \frac{2j\Delta F_{\perp }+l\left( 1-l^{\prime }\right) \Delta
F_{c}}{B_{z}}\right) =  \notag \\
A_{\beta }B_{i,\beta }\left[ 2\cos \left( \frac{4\pi \Delta F_{\perp }}{B_{z}%
}\right) +\sum\limits_{l=\pm 1}\cos \left( 4\pi \frac{\Delta F_{\perp
}+l\Delta F_{c}}{B_{z}}\right) \right] .  \label{3h}
\end{gather}

\begin{figure}[tbh]
\includegraphics[width=0.5\textwidth]{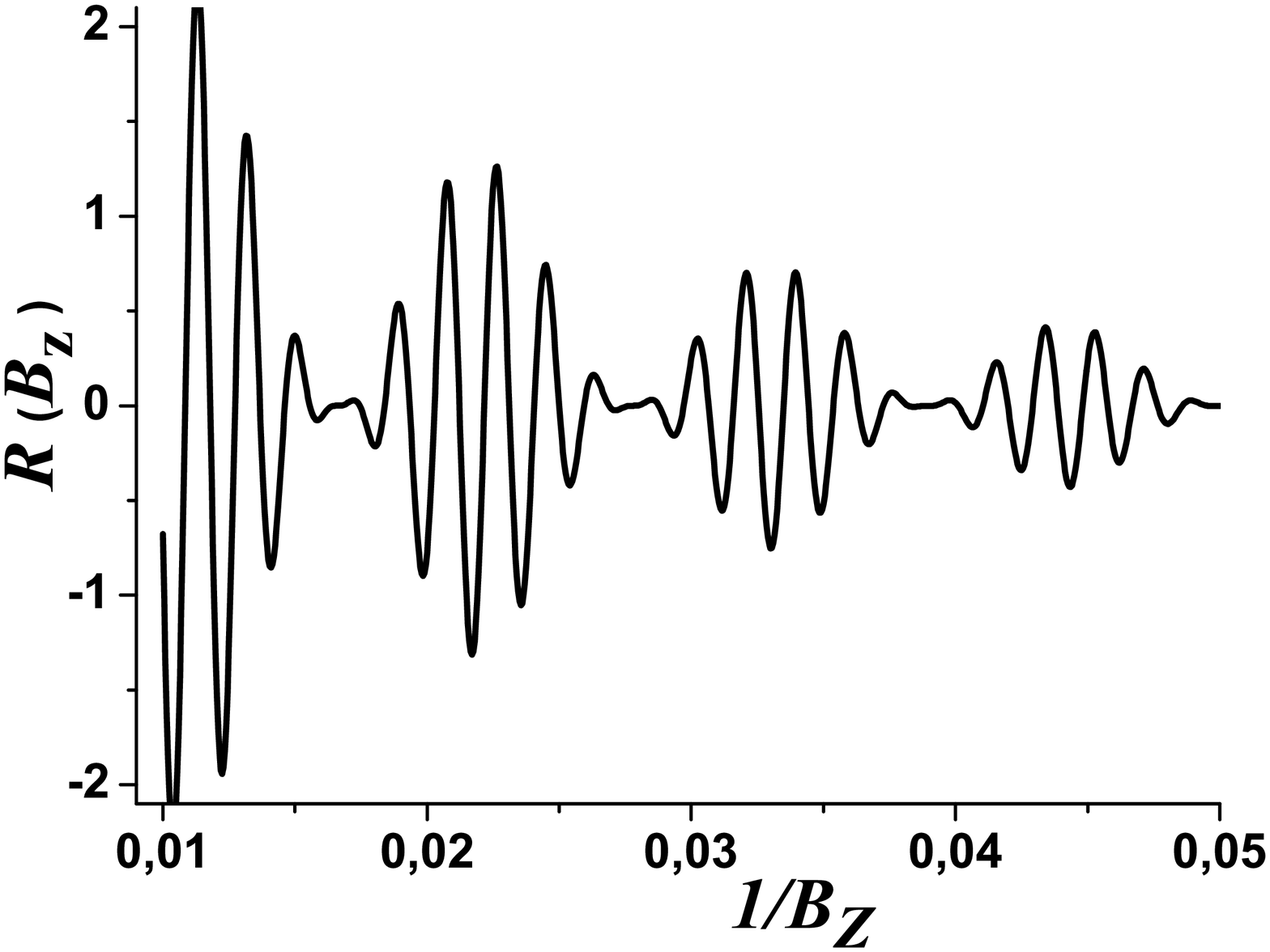} \\
\includegraphics[width=0.5\textwidth]{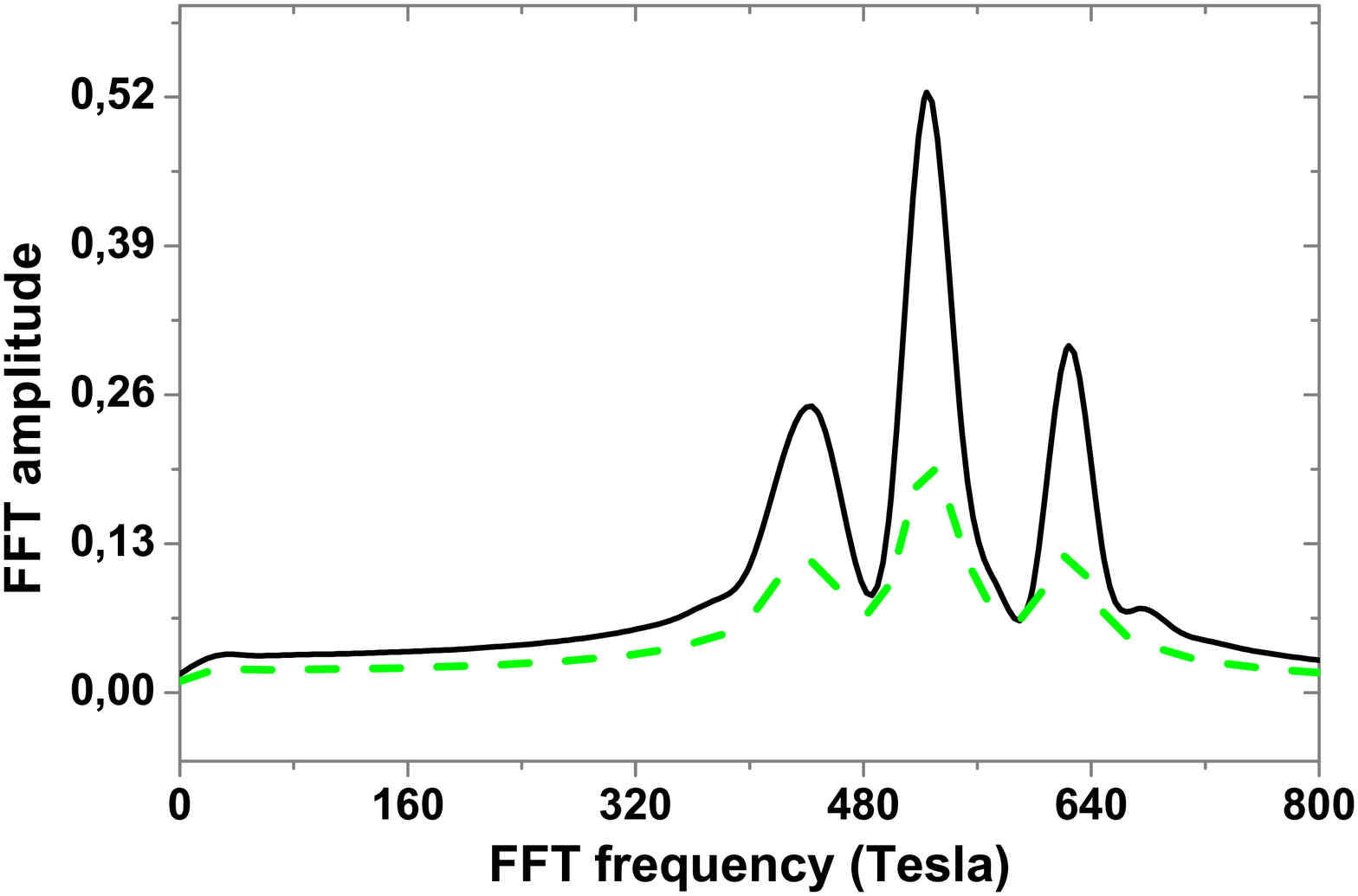}
\caption{The general view (a) and Fourrier transform (b) of SlO of
MR given by Eq. (\protect\ref{3h}) at two different Dingle
temperatures:  $\Gamma /\hbar \protect\omega _{c}(B_{z}=1T)=8$
(solid black line, corresponding to Fig. a) and $16$ (dashed green
line).} \label{Fig2FFT}
\end{figure}

There is an important difference between the SlO with frequency $2\Delta
F_{c}$, produced by FS warping due to $k_{z}$ dispersion from $t_{z}$, and
the SlO frequency $2\Delta F_{\perp }$ from the bilayer splitting $t_{\perp
} $. If due to $t_{z}$, the slow frequency has a strongly non-monotonic
dependence on the tilt angle of the magnetic field $\theta$:\cite%
{SO,CommentDFtz}
\begin{equation}
\Delta F_{c}(\theta )=\Delta F_{c}(\theta =0)J_{0}\left( k_{F}c^{\star }\tan
\theta \right) /\cos \theta ,  \label{Angtz}
\end{equation}%
where $c^{\star }$ ($=11.65\mathring{A}$ for YBCO) is the lattice constant
in the interlayer $z$-direction, and $k_{F}$ is the Fermi momentum. This
angular dependence is as for the beat frequency of MQO\cite{SO} or as for a
square root of interlayer conductivity (AMRO) \cite{KartsAMRO1988,Yagi1990}
in a quasi-2D metal, but differs strongly from the standard cosine dependence%
\begin{equation}
F\left( \theta \right) =F\left( \theta =0\right) /\cos \theta ,
\label{AngBylayer}
\end{equation}%
typical of quasi-2D metals. Eq. (\ref{Angtz}) has an obvious geometrical
interpretation\cite{Yam}: the slightly warped FS has two extremal
cross-sections $S_{ext}$ perpendicular to the magnetic field $\boldsymbol{B}$%
, which become equal, to first order in $t_{z}$, at some tilt angles $\theta
_{Yam}$. The SlO frequency $2F_{\perp }$ due to bilayer splitting, however,
has the cosine angular dependence given by Eq. (\ref{AngBylayer}) rather
than by Eq. (\ref{Angtz}). To show this, consider a magnetic field that is
not very strong, $B\ll F_{\alpha }\approx 530T$, so $\hbar \omega _{c}\ll
4\pi t_{\perp }$ and the field does not modify the electron spectrum of one
bilayer. Consider the dispersion in Eq. (\ref{EpsBilayerKzApprox}),
corresponding to the FS in Fig. \ref{FigFS}. At $t_{z}\ll t_{\perp }$ the FS
consists of two cylinders along the $z$-axis with base areas $S_{even}$ and $%
S_{odd}$, differing by $\Delta S=S_{even}-S_{odd}=4\pi t_{\perp }m^{\ast
}=const$. In a tilted magnetic field the two corresponding extremal
cross-section areas are $S_{even,odd}=S_{even,odd}/\cos \theta $, leading
directly to Eq. (\ref{AngBylayer}). 
The two split FS cylinders are weakly warped at finite interbilayer hopping $%
t_{z}$, 
but if $2t_{z}<t_{\perp }$ and the split FS do not intersect, a finite $%
t_{z} $ does not change the angular dependence in Eq. (\ref{AngBylayer}) to
leading order in $t_{z}/t_{\perp }$.

The amplitude of the central frequency $2\Delta F_{\perp }$\ in Eq. (\ref{3h}%
) is twice as large as the amplitudes of the side frequencies $2\Delta
F_{\perp }\pm 2\Delta F_{c}$, and in Fig. \ref{Fig2FFT} the amplitudes of
side peaks are additionally damped by the finite Dingle factor. This
resembles closely the experimental data in YBCO.\cite%
{AnnuReviewYBCO2015,SebastianRepProgPhys2012,SebastianPRB2010}. We therefore
propose an alternative interpretation of the observed \cite%
{ProustNature2007,SebastianNature2008,AudouardPRL2009,SingletonPRL2010,SebastianPRB2010,SebastianPNAS2010,SebastianPRL2012,SebastianNature2014,ProustNatureComm2015}
oscillations at low frequency $F_{\alpha }\approx 530T$ in YBa$_{2}$Cu$_{3}$O%
$_{6+\delta }$ (and, possibly, in YBa$_{2}$Cu$_{4}$O$_{8}$ \cite%
{Yelland2008,Bangura2008,TanPNAS2015}), in which three equidistant harmonics
are not due to the small pockets 
of the FS reconstructed by CDW order,\cite%
{EfimovPRB2008,GarciaNJP2010,HarrisonNJP2012,HarrisonSciRep2015,Briffa2015,SebastianRepProgPhys2012,ProustComptesRendus2013,AnnuReviewYBCO2015}
but originate rather from mixing, according to Eq. (\ref{3h}), of
four frequencies $F_{\beta }\pm \Delta F_{\perp }\pm \Delta
F_{c}$, formed\ by a fundamental frequency $F_{\beta }$ split by
bilayer and interbilayer electron hopping integrals $t_{\perp }$
and $t_{z}$. In terms of Eqs. (\ref{Slow1})-(\ref{3h}) and
electron dispersion in Eqs. (\ref{EpsBilayerKz}) or
(\ref{EpsBilayerKzApprox}), the frequencies $F_{\alpha }=2\Delta
F_{\perp }\approx 530T=2t_{\perp }B/\hbar \omega _{c}$ and $\Delta
F_{\alpha}=2\Delta F_{c}\approx 90T\approx 4t_{z}B/\hbar \omega
_{c}$ allow us to extract the values of bilayer $t_{\perp }$ and
interbilayer $t_{z}$
average electron transfer integrals from experiment. 
We summarize arguments for this new interpretation:

(1) Experimentally, the multiple extra MQO frequencies predicted from FS
reconstruction are missing\cite%
{EfimovPRB2008,GarciaNJP2010,SebastianPhilTrans2011,HarrisonNJP2012,HarrisonSciRep2015,Briffa2015,ProustComptesRendus2013,AnnuReviewYBCO2015}%
. 
In contrast, in the SlO model
the only frequency with an amplitude $A_{\beta }^{2}$ comparable to the side
peaks at $F_{\alpha }\pm \Delta F_{\alpha}$ would be $2\Delta F_{c}\sim 90T$%
. This low frequency, however, can be detected only at low magnetic field $%
B_{z}<\Delta F_{c}$, where the oscillations are strongly damped by the
Dingle factor. Actually such a frequency may have been seen recently.\cite%
{ProustNatureComm2015} The SlO scenario also predicts a much larger
frequency $F_{\beta }$ from FS pockets, which should be more easily observed
in cyclotron resonance or dHvA than in the Shubnikov-de Haas effect.
Experimentally, the $F_{\beta }\approx 1.65$kT frequency was indeed observed
in dHvA \cite{SebastianNature2008} and Tunnel Diode Oscillation cyclotron
resonance \cite{SebastianPNAS2010} measurements, whereas the $F_{\alpha }$
is much clearer in magnetotransport.

(2) The observed $F_{\alpha }\approx 530T$ depends weakly on the degree of
doping \cite{SingletonPRL2010,SebastianPNAS2010}, more consistent with SlO
than with expectations for small FS-pockets.

(3) The bilayer splitting $t_{\perp }$ expected from band-structure
calculations\cite{Briffa2015,Andersen1995} is consistent with the observed
SlO frequency $F_{\alpha }$: $2\Delta F_{\perp }=2t_{\perp }B/\hbar \omega
_{c}\sim F_{\alpha }=530T\approx 2\%\cdot S_{BZ}$, giving \cite{Commentm} $%
t_{\perp }\equiv \left\langle t_{\perp }\left( \boldsymbol{k}_{\parallel
}\right) \right\rangle =\hbar eF_{\alpha }/2m_{\beta }^{\ast }\approx 8meV$.
Note that the maximum value $t_{\perp }\left( \boldsymbol{k}_{\parallel
}\right) $ of bilayer transfer may considerably exceed this average value $%
t_{\perp }$. Similarly the observed $t_{z}$-induced splitting $2\Delta
F_{c}\approx 90T\sim 4t_{z}B/\hbar \omega _{c}$, gives a reasonable average
value $2t_{z}\approx 1.4meV$.

(4) Long-range spatial inhomogeneities, common in cuprates, should strongly
damp oscillations from FS pockets due to smearing of the Fermi level. They
should affect the proposed SlO much less, similar to slow oscillations in
Ref. \cite{SO}. 

(5) The angular dependence of the observed frequencies\cite{SebastianPRB2010}
corresponds to the SlO interpretation, predicting $F_{\alpha }\left( \theta
\right) =2F_{\perp }\left( \theta \right) \approx const/\cos \theta $ and $%
\Delta F_{c}\cos \theta \propto J_{0}(k_{F}c^{\star }\tan \theta )$, where $%
c^{\star }\approx 11.8\mathring{A}$ is the interlayer lattice constant. The
observed strong angular dependence of the split frequency $\Delta
F_{c}(\theta )$ (see Fig. 4a of Ref. \cite{SebastianPRB2010}) is well fit by
Eq. (\ref{Angtz}), corresponding to the FS-warping origin of this splitting.
The first Yamaji angle $\theta _{Yam}\approx 43^{\circ }$ in $\Delta
F_{c}\left( \theta \right) $, corresponding to the first zero of the Bessel
function $J_{0}\left( k_{F}c^{\star }\tan \theta \right) $ in Eq. (\ref%
{Angtz}), is clearly seen in Fig. 4a of Ref. \cite{SebastianPRB2010}. This
Yamaji angle gives the Fermi momentum $k_{F}=2.4/c^{\star }\tan \theta
_{Yam} = 2.2nm^{-1}$ and FS-pocket area of about $S_{ext}\sim \pi
k_{F}^{2}\approx 15nm^{-2}$, corresponding to MQO frequency $%
F_{0}=S_{ext}\hbar /2\pi e\approx 1.6$kT, which is far from $F_{\alpha
}\approx 530$T but close to $F_{\beta }\approx 1.65$kT observed in \cite%
{SebastianNature2008,SebastianPNAS2010}. Thus we conclude that the FS pocket
responsible for the observed oscillations, has an area corresponding to $%
F_{\beta }$ rather than $F_{\alpha }$, which is a slow frequency
corresponding to $t_{\perp }$.
Note that $F_{\beta }$ was observed only at rather low temperatures and not
in all samples,\cite{SebastianPhilTrans2011} consistent with our model
because frequencies corresponding to real FS pockets are more strongly
damped by temperature than the proposed SlO and strongly damped by
sample-dependent inhomogeneities.

(6) An underlying CDW is not a necessary prerequisite for
observation of SlO. There are no issues to be resolved as to how a
weak and fluctuating CDW ordering could overcome magnetic
breakdown, which should be strong for fields up to 100 tesla.

(7) The relative amplitudes of the frequencies are naturally explained
without additional fitting parameters (see Fig. \ref{Fig2FFT}).

We note possible counter-arguments:

(1) The observed magnetoresistance oscillations are quite strongly damped by
temperature, with effective mass $m^{\star }\approx 1.6m_{e}$,\cite%
{SebastianPRB2010,SingletonPRL2010} with $m_{e}$ the free electron mass, in
disagreement with the simple form predicted here\cite{SO,Shub} but may arise
from the square of the Dingle factor $R_{D}$ with temperature dependence
enhanced by electron-phonon and the very strong electron-electron
interaction in cuprates.%

(2) A component at $F_{\alpha }$ is also observed in the dHvA effect.
This can be due to the electron-electron interaction, roughly proportional
to the product of the density of states, giving a nonlinearity in the
magnetization as a functional of the oscillating density of states.

(3) Oscillations with a frequency 
$\approx 840T$ (but without side-frequency splitting), corresponding to 3\%
of the Brillouin zone, have been observed in HgBa$_{2}$CuO$_{4+\delta }$,
where there is no bilayer splitting, but a much larger $t_{z}$ is expected
from the shorter interlayer distance. In fact this is consistent with SlO,
but as the spectrum is simpler the evidence is not as compelling.


To summarize, we propose an alternative interpretation of the observed
magnetic oscillations in YBa$_{2}$Cu$_{3}$O$_{6+\delta }$ high-Tc
superconductors, with frequencies $F_{\alpha }\approx 530T$ and $F_{\alpha
}\pm \Delta F_{\alpha}$ related to the bilayer splitting and corrugation
rather than to tiny FS pockets. This is based on the new result (\ref{3h}%
), as illustrated in Fig. \ref{Fig2FFT}. The frequencies allow us to
estimate values of bilayer splitting $t_{\perp }=\hbar eF_{\alpha }/2m^{\ast
}$ and of $k_{z}$-dispersion $t_{z}=\hbar e\Delta F_{\alpha}/4m^{\ast }$.
Angular dependence points to a true FS pocket close to $F_{\beta}\approx 1.6
kT$. While this can explain current observations without recourse to CDWs,
neither does it rule out their potential to make new frequencies appear.
Such frequencies would, however, be more sensitive to inhomogeneity as well
as magnetic breakdown.

\acknowledgments{The authors thank Ted Forgan for raising this
issue and communication of unpublished results. The work is
supported by the Ministry of Education and Science of the Russian Federation in the framework of Increase Competitiveness Program of MISiS. P.G. acknowledges the Russian Science Foundation grant \# 16-42-01100.}


\begin{thebibliography}{99}
\bibitem{Abrik} A.A. Abrikosov, \textit{Fundamentals of the theory of metals}%
, North-Holland, 1988.

\bibitem{Shoenberg} Shoenberg D. \textit{\ Magnetic oscillations in metals},
Cambridge University Press 1984.

\bibitem{Ziman} J. M. Ziman, \textit{Principles of the Theory of Solids},
Cambridge Univ. Press 1972.

\bibitem{ProustNature2007} Nicolas Doiron-Leyraud, Cyril Proust, David
LeBoeuf, Julien Levallois, Jean-Baptiste Bonnemaison, Ruixing Liang, D. A.
Bonn, W. N. Hardy, Louis Taillefer, Nature \textbf{447}, 565 (2007).

\bibitem{AnnuReviewYBCO2015} Suchitra E. Sebastian and Cyril Proust, Annu.
Rev. Condens.Matter Phys. \textbf{6}, 411 (2015) and references therein.

\bibitem{HelmNd2015} T. Helm, M. V. Kartsovnik, C. Proust, B. Vignolle, C.
Putzke, E. Kampert, I. Sheikin, E.-S. Choi, J. S. Brooks, N. Bittner, W.
Biberacher, A. Erb, J. Wosnitza, and R. Gross, Phys. Rev. B \textbf{92},
094501 (2015) and references therein.

\bibitem{BaFeAs2011} Taichi Terashima, Nobuyuki Kurita, Megumi Tomita,
Kunihiro Kihou, Chul-Ho Lee, Yasuhide Tomioka, Toshimitsu Ito, Akira Iyo,
Hiroshi Eisaki, Tian Liang, Masamichi Nakajima, Shigeyuki Ishida, Shin-ichi
Uchida, Hisatomo Harima, and Shinya Uji, Phys. Rev. Lett. \textbf{107},
176402 (2011).

\bibitem{Graf2012} D. Graf, R. Stillwell, T. P. Murphy, J.-H. Park, E. C.
Palm, P. Schlottmann, R. D. McDonald, J. G. Analytis, I. R. Fisher, and S.
W. Tozer, Phys. Rev. B. \textbf{85}, 134503 (2012).

\bibitem{ColdeaReview2013} Amalia I. Coldea, Daniel Braithwaite, Antony
Carrington, Comptes Rendus Physique \textbf{14}, 94 (2013).

\bibitem{FeSeTerashima2014} T. Terashima, N. Kikugawa, A. Kiswandhi, E.-S.
Choi, J. S. Brooks, S. Kasahara, T. Watashige, H. Ikeda, T. Shibauchi, Y.
Matsuda, T. Wolf, A. E. B\"{o}hmer, F. Hardy, C. Meingast, H. v. L\"{o}%
hneysen, M.-T. Suzuki, R. Arita, and S. Uji, Phys. Rev. B. \textbf{90},
144517 (2014).

\bibitem{FeSeAudouard2015} A. Audouard, F. Duc, L. Drigo, P. Toulemonde, S.
Karlsson, P. Strobel, and A. Sulpice, Europhys. Lett. \textbf{109}, 27003
(2015).

\bibitem{FeSeWatsonPRB2015} M. D. Watson, T. K. Kim, A. A. Haghighirad, N.
R. Davies, A. McCollam, A. Narayanan, S. F. Blake, Y. L. Chen, S.
Ghannadzadeh, A. J. Schofield, M. Hoesch, C. Meingast, T. Wolf, and A. I.
Coldea, Phys. Rev. B \textbf{91}, 155106 (2015).

\bibitem{FeSeMQOPRL2015} M.\thinspace D. Watson, T. Yamashita, S. Kasahara,
W. Knafo, M. Nardone, J. Beard, F. Hardy, A. McCollam, A. Narayanan,
S.\thinspace F. Blake, T. Wolf, A.\thinspace A. Haghighirad, C. Meingast,
A.\thinspace J. Schofield, H. v. L\"{o}hneysen, Y. Matsuda, A.\thinspace I.
Coldea, and T. Shibauchi, Phys. Rev. Lett. \textbf{115}, 027006 (2015).

\bibitem{SebastianNature2008} Suchitra E. Sebastian, N. Harrison, E. Palm,
T. P. Murphy, C. H. Mielke, Ruixing Liang, D. A. Bonn, W. N. Hardy and G. G.
Lonzarich, Nature 454, 200 (2008)

\bibitem{AudouardPRL2009} Alain Audouard, Cyril Jaudet, David Vignolles,
Ruixing Liang, D. A. Bonn, W. N. Hardy, Louis Taillefer, and Cyril Proust,
Phys. Rev. Lett. \textbf{103}, 157003 (2009).

\bibitem{SingletonPRL2010} John Singleton, Clarina de la Cruz, R. D.
McDonald, Shiliang Li, Moaz Altarawneh, Paul Goddard, Isabel Franke, Dwight
Rickel, C. H. Mielke, Xin Yao, and Pengcheng Dai, Phys. Rev. Lett. \textbf{%
104}, 086403 (2010).

\bibitem{SebastianPNAS2010} S. E. Sebastian, N. Harrison, M. M. Altarawneh,
C. H. Mielke, R. Liang, D. A. Bonn, W. N. Hardy, G. G. Lonzarich, Proc.
Natl. Acad. Sci. U.S.A. \textbf{107}, 6175 (2010).

\bibitem{SebastianPRB2010} Suchitra E. Sebastian, N. Harrison, P. A.
Goddard, M. M. Altarawneh, C. H. Mielke, Ruixing Liang, D. A. Bonn, W. N.
Hardy, O. K. Andersen, and G. G. Lonzarich, Phys. Rev. B \textbf{81}, 214524
(2010).

\bibitem{SebastianPRL2012} Suchitra E. Sebastian, N. Harrison, Ruixing
Liang, D. A. Bonn, W. N. Hardy, C. H. Mielke, and G. G. Lonzarich, Phys.
Rev. Lett. \textbf{108}, 196403 (2012).

\bibitem{SebastianNature2014} S.E. Sebastian, N. Harrison, F.F. Balakirev,
M.M. Altarawneh, P.A. Goddard et al., Nature \textbf{511}, 61 (2014).

\bibitem{ProustNatureComm2015} N. Doiron-Leyraud, S. Badoux, S. Rene de
Cotret, S. Lepault, D. LeBoeuf, F. Laliberte, E. Hassinger, B.J. Ramshaw,
D.A. Bonn, W.N. Hardy,R. Liang, J.-H.. Park, D. Vignolles, B. Vignolle, L.
Taillefer and C. Proust, Nature Comm. \textbf{6}, 6034 (2015).

\bibitem{SebastianRepProgPhys2012} S. E. Sebastian, N. Harrison, and G. G.
Lonzarich, Rep. Prog. Phys. \textbf{75}, 102501 (2012).

\bibitem{ProustComptesRendus2013} Baptiste Vignolle, David Vignolles,
Marc-Henri Julien, and Cyril Proust, Comptes Rendus Phys. \textbf{14}, 39
(2013).

\bibitem{SebastianPhilTrans2011} Suchitra E. Sebastian, Neil Harrison and
Gilbert G. Lonzarich, Phil. Trans. R. Soc. A \textbf{369}, 1687 (2011).

\bibitem{EfimovPRB2008} I. S. Elfimov, G. A. Sawatzky, and A. Damascelli,
Phys. Rev. B \textbf{77}, 060504(R) (2008).

\bibitem{GarciaNJP2010} David Garcia-Aldea and Sudip Chakravarty, New
Journal of Physics \textbf{12}, 105005 (2010).

\bibitem{HarrisonNJP2012} N. Harrison and S. E. Sebastian, New Journal of
Physics \textbf{14}, 095023 (2012).

\bibitem{HarrisonSciRep2015} N. Harrison, B. J. Ramshaw and A. Shekhter,
Scientific Reports \textbf{5}, 10914 (2015).

\bibitem{Briffa2015} A. K. R. Briffa, E. Blackburn, S. M. Hayden, E. A.
Yelland, M. W. Long, and E. M. Forgan, Phys. Rev. B \textbf{93}, 094502
(2016).

\bibitem{XRayScience2012} G. Ghiringhelli, M. Le Tacon, M. Minola, S.
Blanco-Canosa, C. Mazzoli, N. Brookes, G. De Luca, A. Frano, D. Hawthorn, F.
He et al., Science \textbf{337}, 821 (2012).

\bibitem{XRayNatPhys2012} J. Chang, E. Blackburn, A. Holmes, N. Christensen,
J. Larsen, J. Mesot, R. Liang, D. Bonn, W. Hardy, A. Watenphul et al., Nat.
Phys. \textbf{8}, 871 (2012).

\bibitem{XRayPRL2012} A. J. Achkar, R. Sutarto, X. Mao, F. He, A. Frano, S.
Blanco-Canosa, M. Le Tacon, G. Ghiringhelli, L. Braicovich, M. Minola, M.
Moretti Sala, C. Mazzoli, R. Liang, D. A. Bonn, W. N. Hardy, B. Keimer, G.
A. Sawatzky, and D. G. Hawthorn, Phys. Rev. Lett. \textbf{109}, 167001
(2012).

\bibitem{Xray2016} S. Gerber, H. Jang, H. Nojiri, S. Matsuzawa, H. Yasumura,
D. A. Bonn, R. Liang, W. N. Hardy, Z. Islam, A. Mehta, S. Song, M. Sikorski,
D. Stefanescu, Y. Feng, S. A. Kivelson, T. P. Devereaux, Z.-X. Shen, C.-C.
Kao, W.-S. Lee, D. Zhu, J.-S. Lee, Science \textbf{350}, 949 (2015); H.
Jang, W.-S. Lee, H. Nojiri, S. Matsuzawa, H. Yasumura, L. Nie, A. V.
Maharaj, S. Gerber, Y. Liu, A. Mehta, D. A. Bonn, R. Liang, W. N. Hardy, C.
A. Burns, Z. Islam, S. Song, J. Hastings, T. P. Devereaux, Z.-X. Shen, S. A.
Kivelson, C.-C. Kao, D. Zhu, J.-S. Lee, arXiv:1607.05359.

\bibitem{NMR2011Wu} T. Wu, H. Mayaffre, S. Kr{\" a}mer, M. Horvatic, C.
Berthier, W. Hardy, R. Liang, D. Bonn, and M.-H. Julien, Nature (London)
\textbf{477}, 191 (2011).

\bibitem{NMR2015Wu} T. Wu, H. Mayaffre, S. Kr{\" a}amer, M. Horvatic, C.
Berthier, W. Hardy, R. Liang, D. Bonn, and M.-H. Julien, Nat. Commun.
\textbf{6}, 6438 (2015).

\bibitem{CDWSoundVelocity} David LeBoeuf, S. Kr{\" a}mer, W. N. Hardy,
Ruixing Liang, D. A. Bonn \& Cyril Proust, Nature Physics \textbf{9}, 79
(2013).

\bibitem{Yelland2008} E. A. Yelland, J. Singleton, C. H. Mielke, N.
Harrison, F. F. Balakirev, B. Dabrowski, and J. R. Cooper, Phys. Rev. Lett.
\textbf{100}, 047003 (2008).

\bibitem{Bangura2008} A. F. Bangura, J. D. Fletcher, A. Carrington, J.
Levallois, M. Nardone, B. Vignolle, P. J. Heard, N. Doiron-Leyraud, D.
LeBoeuf, L. Taillefer, S. Adachi, C. Proust, and N. E. Hussey, Phys. Rev.
Lett. \textbf{100}, 047004 (2008).

\bibitem{TanPNAS2015} B. S. Tan, N. Harrison, Z. Zhu, F. Balakirev, B. J.
Ramshaw, A. Srivastava, S. A. Sabok-Sayr, B. Dabrowski, G. G. Lonzarich, and
Suchitra E. Sebastian, Proc. Natl. Acad. Sci. U.S.A. \textbf{112}, 9568
(2015).

\bibitem{CommentDoping1} Doping changes the size of unreconstructed FS
considerably , giving a very large relative change of the size of any small
reconstructed FS pockets.

\bibitem{HelmNd2009} T. Helm, M.V. Kartsovnik, M. Bartkowiak, N. Bittner, M.
Lambacher, A. Erb, J. Wosnitza, and R. Gross, Phys. Rev. Lett. \textbf{103},
157002 (2009).

\bibitem{HelmNd2010} T. Helm, M.V. Kartsovnik, I. Sheikin, M. Bartkowiak, F.
Wolff-Fabris, N. Bittner, W. Biberacher, M. Lambacher, A. Erb, J. Wosnitza,
and R. Gross, Phys. Rev. Lett. \textbf{105}, 247002 (2010).

\bibitem{DopingDependence2015} B. J. Ramshaw, S. E. Sebastian, R. D.
McDonald, James Day, B. S. Tan, Z. Zhu, J. B. Betts, Ruixing Liang, D. A.
Bonn, W. N. Hardy, N. Harrison, Science \textbf{348}, 317 (2015). See in
particular Fig. 2.

\bibitem{NPhysPereg} T. Pereg-Barnea, H.Weber, G. Refael and M.
Franz, Quantum oscillations from Fermi arcs. Nat. Phys.
\textbf{6}, 44-49 (2010).


\bibitem{FournierARPES2010} D. Fournier, G. Levy, Y. Pennec, J. L.
McChesney, A. Bostwick, E. Rotenberg, R. Liang, W. N. Hardy, D. A. Bonn, I.
S. Elfimov \& A. Damascelli, Nature Physics \textbf{6}, 905 (2010).

\bibitem{ElPocketHall2007} David LeBoeuf, Nicolas Doiron-Leyraud, Julien
Levallois, R. Daou, J.-B. Bonnemaison, N. E. Hussey, L. Balicas, B. J.
Ramshaw, Ruixing Liang, D. A. Bonn, W. N. Hardy, S. Adachi, Cyril Proust \&
Louis Taillefer, Nature \textbf{450}, 533 (2007).

\bibitem{BadouxHall2016} S. Badoux, W. Tabis, F. Laliberte, G.
Grissonnanche, B. Vignolle, D. Vignolles, J. Beard, D. A. Bonn, W. N. Hardy,
R. Liang, N. Doiron-Leyraud, Louis Taillefer \& Cyril Proust, Nature \textbf{%
531}, 210 (2016).

\bibitem{ElPocketSeebeck2010} J. Chang, R. Daou, Cyril Proust, David
LeBoeuf, Nicolas Doiron-Leyraud, Francis Laliberte, B. Pingault, B. J.
Ramshaw, Ruixing Liang, D. A. Bonn, W. N. Hardy, H. Takagi, A. B. Antunes,
I. Sheikin, K. Behnia, and Louis Taillefer, Phys. Rev. Lett. \textbf{104},
057005 (2010).

\bibitem{SO} M.V. Kartsovnik, P.D. Grigoriev, W. Biberacher, N.D. Kushch, P.
Wyder, Phys. Rev. Lett. \textbf{89}, 126802 (2002).

\bibitem{Shub} P.D. Grigoriev, Phys. Rev. B \textbf{67}, 144401 (2003)
[arXiv:cond-mat/0204270].

\bibitem{RET} P.D. Grigoriev, A. A. Sinchenko, P. Lejay, A. Hadj-Azzem, J.
Balay, O. Leynaud, V. N. Zverev and P. Monceau, Eur. Phys. J. B \textbf{89},
151 (2016). In this case slow oscillations in bilayer rare-earth tellurides
at a single frequency were interpreted as coming from an interlayer coupling
but with no warping.

\bibitem{Dingle} R.B. Dingle, Proc. Roy. Soc. \textbf{A211,} 517 (1952).

\bibitem{Bychkov} Yu. A. Bychkov, Zh. Exp. Theor. Phys. \textbf{39}, 1401
(1960), [Sov. Phys. JETP \textbf{12}, 977 (1961)].

\bibitem{CommentMB} The amplitudes $A_{\beta },B_{i,\beta }$ may also
contain the magnetic-breakdown damping factors\cite{Shoenberg} if there are
FS pockets close in momentum space, as is the case for FS reconstruction by
a CDW.

\bibitem{Champel2001} V. M. Gvozdikov, Fiz. Tverd. Tela (Leningrad) 26, 2574
(1984) [Sov. Phys. Solid State 26, 1560 (1984)]; T. Champel and V. P.
Mineev, Phil. Magazine B \textbf{81}, 55 (2001).

\bibitem{CommentDFtz} Eq. (\ref{Angtz}) assumes that $t_{z}\left(
\boldsymbol{k}_{\parallel }\right) \approx const$. When $t_{z}\left(
\boldsymbol{k}_{\parallel }\right) $ has a strong angular dependence, Eq. (%
\ref{Angtz}) is modified,\cite{Bergemann,GrigAMRO2010} but remains an
oscillating function of $k_{F}c^{\star }\tan \theta $.

\bibitem{KartsAMRO1988} M.V. Kartsovnik, P. A. Kononovich , V. N. Laukhin
and I. F. Shchegolev, JETP Lett. \textbf{48}, 541 (1988).

\bibitem{Yagi1990} R. Yagi, Y. Iye, T. Osada, S. Kagoshima, J. Phys. Soc.
Jpn. \textbf{59}, 3069 (1990).

\bibitem{Yam} K. Yamaji, J. Phys. Soc. Jpn. \textbf{58}, 1520 (1989).


\bibitem{Andersen1995} O. K. Andersen, A.I. Liechtenstein, O. Jepsen, F.
Paulsen, J. Phys. Chem. Solids 56, 1573 (1995).

\bibitem{Commentm} We take $m_{\beta }^{\ast }\approx 3.8m_{e}$ as obtained
from dHvA measurements for $\beta $-frequency,\cite{SebastianNature2008}
because within our model $m_{\alpha }^{\ast }\approx 1.6m_{e}$ is not a true
effective electron mass but rather an effective parameter coming from the
temperature dependence of the square of Dingle temperature.

\bibitem{Bergemann} C. Bergemann, S. R. Julian, A. P. Mackenzie, S.
NishiZaki, and Y. Maeno, Phys. Rev. Lett. \textbf{84}, 2662 (2000).

\bibitem{GrigAMRO2010} P.D. Grigoriev, Phys. Rev. B \textbf{81}, 205122
(2010).
\end{thebibliography}
\end{document}